\documentclass[twocolumn,secnumarabic,amssymb, nobibnotes, prx,superscriptaddress]{revtex4-2} 
\usepackage{CJK}
\usepackage{graphicx} 
\usepackage{blindtext}
\usepackage{physics} 

\usepackage{lipsum}
\usepackage{amsfonts,amssymb,amsmath}
\usepackage{bm}
\usepackage{xcolor,calc}
\usepackage{ulem}
\usepackage{hyperref}
\usepackage{siunitx}

\newcommand{\gev}{GeV }
\newcommand{\gevv}{GeV}

\newcommand{\nuc}{${}^{13}$C }
\newcommand{\nucc}{${}^{13}$C}

\newcommand{\Tone}{$T_{1,e}$ }
\newcommand{\Ttwo}[1]{$T_{2,{{#1}}}$}
\newcommand{\Ttwostar}[1][]{$T_{2,#1}^{\star}$}
\newcommand{\taudd}{$\tau_{\text{DD}}$ }

\begin{document}


\title{High-Fidelity Single-Shot Readout and Selective Nuclear Spin Control for a Spin-1/2 Quantum Register in Diamond}

\author{Prithvi Gundlapalli}
\email{prithvi.gundlapalli@alumni.uni-ulm.de}
\affiliation{Institute for Quantum Optics, Ulm University, Albert-Einstein-Allee 11, D-89081 Ulm, Germany}
\author{Philipp J. Vetter}
\affiliation{Institute for Quantum Optics, Ulm University, Albert-Einstein-Allee 11, D-89081 Ulm, Germany}
\author{Genko Genov}
\affiliation{Institute for Quantum Optics, Ulm University, Albert-Einstein-Allee 11, D-89081 Ulm, Germany}
\author{Michael Olney-Fraser}
\affiliation{Institute for Quantum Optics, Ulm University, Albert-Einstein-Allee 11, D-89081 Ulm, Germany}
\author{Peng Wang}
\affiliation{Institute for Quantum Optics, Ulm University, Albert-Einstein-Allee 11, D-89081 Ulm, Germany} 
\author{Matthias M. M\"uller}
\affiliation{Peter Gr\"unberg Institute-Quantum Control (PGI-8), Forschungszentrum J\"ulich GmbH, D-52425 J\"ulich, Germany}
\author{Katharina Senkalla}
\affiliation{Institute for Quantum Optics, Ulm University, Albert-Einstein-Allee 11, D-89081 Ulm, Germany}
\author{Fedor Jelezko}
\email{fedor.jelezko@uni-ulm.de}
\affiliation{Institute for Quantum Optics, Ulm University, Albert-Einstein-Allee 11, D-89081 Ulm, Germany}
\date{\today}

\begin{abstract}
    Quantum networks offer a way to overcome the size and complexity limitations of single quantum devices by linking multiple nodes into a scalable architecture.
    Group-IV color centers in diamond, paired with long-lived nuclear spins, have emerged as promising building blocks demonstrating proof-of-concept experiments such as blind quantum computing and quantum-enhanced sensing.
    However, realizing a large-scale electro-nuclear register remains a major challenge.
    Here we establish the germanium-vacancy (\gevv) center as a viable platform for such network nodes.
    Using correlation spectroscopy, we identify single nuclear spins within a convoluted spin environment, overcoming limitations imposed by the color center's spin-$1/2$ nature and thereby enabling indirect control of these nuclear spins.
    We further demonstrate high-fidelity single-shot readout of both the \gev center ($95.8\,\%$) and a neighboring \nuc nuclear spin ($93.7\,\%$), a key tool for feed-forward error correction.
    These critical advances position the \gev center as a compelling candidate for next-generation quantum network nodes.
\end{abstract}

\maketitle


\section{Introduction}
Distributing entanglement across large distances transforms isolated quantum devices into powerful networks, advancing quantum communication, computation, and sensing~\cite{azuma2023quantum}.
Prototypes of quantum networks based on group-IV color centers in diamond have already extended to metropolitan scales~\cite{knaut2024entanglement}, demonstrating concepts such as blind quantum computing~\cite{wei2025universal} and non-local sensing~\cite{stas2025entanglement}. 
Despite these achievements, current implementations remain constrained by the small number of accessible qubits per node. 
While surrounding nuclear spins could provide a range of functionalities such as acting as long-lived resources for information storage, deterministic entanglement, and error correction~\cite{taminiau2014universal,grimm2025coherent,wei2025universal}, their precise identification and control remains challenging due to spin-$1/2$ nature of these color centers.
Typically, surrounding nuclear spins are identified by acquiring a spectrum of the spin environment via dynamical decoupling (DD)~\cite{taminiau2012detection, klotz2025bipartite, nguyen2019integrated}. 
For spin-$1/2$ systems, such sequences provide only second-order sensitivity to the interaction with the nuclear spins, causing overlapping resonances for distant ones~\cite{zahedian24blueprint,beukers2025control,nguyen2019integrated}.
Moreover, interaction with strongly-coupled nearby nuclear spins can cause additional resonances, masking the signals of more weakly-coupled spins~\cite{casanova2015robust, zahedian24blueprint}. \\
In this work, we extend the method of correlation spectroscopy (CS) \cite{laraoui2013high,Kong2015prapplied,Staudacher2015natcommun} to two dimensions to overcome these problems, facilitating the detection of distant, weakly-coupled \nuc nuclear spins in a convoluted spin environment surrounding a negatively-charged germanium-vacancy (\gevv) center. 
We further showcase how the sequence can identify resonances in the DD spectrum corresponding to single nuclear spins, enabling their selective control via the \gev center.
This enables the implementation of conditional gates which we use for measurement-based initialization (MBI) of a weakly-coupled nuclear spin.
Such MBI requires high-fidelity readout of the associated color center, a capability also essential for tasks such as error correction that depend on feed-forward operations~\cite{abobeih2022fault,mallet2009single,salathe2018low,xiang2020simultaneous}. 
We achieve a record-high single-shot readout (SSR) fidelity of  $95.80\,\%$ on our \gev center enabled by composite pulses that mitigate state preparation errors and an anti-correlation check that rejects blinking events.
Additionally, we demonstrate SSR on a nearby \nuc nuclear spin using optimized pulses, yielding a fidelity of $93.66\,\%$. \\
Our CS findings are applicable to any spin-$1/2$ system. The demonstration of high-fidelity readout and efficient control of an electro–nuclear multi-qubit spin register in this work paves the way for scalable quantum network nodes based on group-IV color centers in diamond.
%

\begin{figure*}
    \centering
    \includegraphics{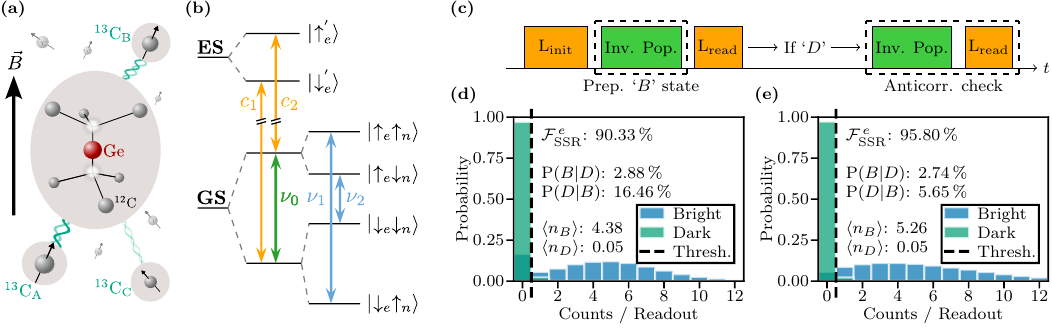}
        \caption{
        Level scheme and single-shot readout (SSR) of the \gev center.
        \quad(a) Physical structure of the \gev center and its coupling to surrounding nuclear spins.
        (b) Reduced energy level scheme of the \gev center showing the hyperfine splitting induced by a strongly-coupled \nuc nuclear spin.
        %
        %
        %
        (c) SSR pulse sequence with additional anti-correlation check.
        The dark state ($\ket{\downarrow_e}$, `$D$') is prepared via a laser pulse on $c_2$ while the bright state ($\ket{\uparrow_e}$, `$B$') is prepared with an additional strong microwave $\pi$ pulse on $\nu_0$ (Inv. Pop.).
        The anti-correlation check is done using an additional population inversion for a `D' outcome. 
        If the subsequent readout is not `B', the event is classified as blinking and discarded. 
        %
        %
        %
        (d) Photon count distribution for the `$D$' state (green) and `$B$' state (blue), yielding a single-shot readout fidelity of $90.33\%$.
        (e) Improved SSR of the \gev center using an anti-correlation check for `$D$' outcomes and composite pulses for improving `$B$' state preparation fidelity, leading to an SSR fidelity of $\mathcal{F}_{\text{SSR}}^{\,e}=95.80\,\%$.
        %
        }
    \label{fig:fig1}
\end{figure*}

\section{Experimental Description}
The experiments are performed on a  $\langle1,1,1\rangle$-oriented, high-pressure high-temperature (HPHT) grown diamond with naturally-abundant \nuc \cite{palyanov2015germanium} in a dilution refrigerator operating at a base temperature of $\sim\,80\,\text{mK}$ \cite{senkalla2024germanium,grimm2025coherent}.
The system features a single, negatively-charged \gev center coupled to surrounding \nuc spins (\autoref{fig:fig1}\,(a)), embedded in a $10\,\text{\textmu m}$ diameter solid immersion lens (SIL).
Applying an external magnetic field of $B_z\approx97\,\text{mT}$ along the principal axis of the \gev center gives access to the electron spin qubit manifold which can be coherently controlled via microwave fields.
Optical addressing and readout are performed using a tunable laser resonant on optical transitions $c_{1,2}$, enabling an initialization fidelity of $98\,\%$~\cite{senkalla2024germanium,grimm2025coherent}.
The coupling of the \gev to the surrounding \nuc is indicated by wavy lines in \autoref{fig:fig1}\,(a).
The strong coupling to a nearby \nuc leads to further splitting of the electron spin sublevels~\cite{grimm2025coherent} as shown in \autoref{fig:fig1}\,(b).
Further details about the setup can be found in~\cite{supplement}. \\
%

\section{Single-Shot Readout on the GeV Center}
An essential parameter describing the feasibility of single-shot readout (SSR) and quantum non-demolition (QND) measurements of the \gev center is the cyclicity of the optical transitions.
Cyclicity is defined as the ratio of spin-conserving to spin-flipping transition rates, indicating the average number of photons emitted during a readout before a spin flip occurs \cite{rosenthal2024single,sukachev2017silicon}.
We estimate a cyclicity of $\sim10^4$, enabling SSR despite the system's low collection efficiency of $\sim0.09\%$~\cite{supplement}.
The $c_2$ transition is used for electron spin initialization and readout, with $\ket{\downarrow_e}$ denoted as the `dark' (`$D$') state and $\ket{\uparrow_e}$ as the `bright' (`$B$') state.
State discrimination is achieved by counting photons detected in the phonon sideband~\cite{senkalla2024germanium} during a short laser pulse whose duration is optimized to maximize the fidelity.
\autoref{fig:fig1}\,(b) shows the photon counts per readout for the bright state (blue) and the dark state (green) during the readout laser pulse.
On average, we observe $\langle{n}_B\rangle=4.38$ and $\langle{n}_D\rangle=0.05$ photons for the bright and dark states respectively.
%
The average SSR fidelity is given by~\cite{robledo2011high}
\begin{equation}
    \mathcal{F}_{\text{SSR}} = 1 - \frac{P( B\vert D)}{2}- \frac{P(D \vert B)}{2}
    \label{eq:ssr_fidelity}
\end{equation}
where $P(i \vert j)$ is the conditional probability of the SSR yielding $i$ when the input state was $j$.
With a readout laser duration of $60\,\text{\textmu s}$ we achieve $\mathcal{F}_{\text{SSR}}^{\,e}=90.33\,\%$.
The high $P(D \vert B) = 16.46\,\%$ is the dominant source of SSR infidelity and is primarily caused by state preparation errors and blinking of the \gev center.\\
The bright state is prepared by first initializing the system into the dark state with a $5\,\text{ms}$ laser pulse followed by a rectangular microwave $\pi$ pulse on transition $\nu_0$ (i.e. the average of $\nu_1$ and $\nu_2$) in \autoref{fig:fig1}\,(b).
Due to strong hyperfine interaction with a nearby \nuc spin, the microwave transitions $\nu_1$ and $\nu_2$ depicted in \autoref{fig:fig1},(b) (green lines) are separated by $|A_{zz}|=(2\pi)\,2.963\,\text{MHz}$.
Combined with the Rabi drive of $\Omega_e = (2\pi)\,7.65\,\text{MHz}$ which is constrained by the setup's cooling power, this interaction significantly reduces the $\pi$ pulse fidelity to $96.4\,\%$ as estimated by numerical simulations.
Therefore, we explore analytic composite pulses which are known to be robust to such an effective detuning~\cite{genov2020universal}.
Our simulations show that the 5-composite-pulse sequence U5b~\cite{genov14correction,genov2020universal} can achieve a population transfer of $99.3\,\%$ with a lower Rabi frequency of $\Omega_e = (2\pi)\,4\,\text{MHz}$ which has a similar fluence~\cite{vetter2024gate} as the single $\Omega_e = (2\pi)\,7.65\,\text{MHz}$ pulse. \\
Blinking, the other major factor limiting SSR fidelity, describes a phenomenon in which the color center enters a state in which no fluorescence for either optical transition $c_{1,2}$ can be detected.
This has been observed for various group-IV color centers and is mostly attributed to charge-state changes \cite{wang2024transform,rosenthal2024single,gorlitz2022coherence,gali13abinitio,chen2025optical,bushmakin2024two,ikeda2025charge}.
In our experiments, blinking manifests as a drop to $\sim 0$ detected photons per readout, a state in which the \gev center can remain for several seconds, see Methods. 
Unlike the bright state SSR fidelity $P( B\vert B)$, the dark state SSR fidelity $P( D\vert D)$ is biased higher by blinking events as the dark spin state is indistinguishable from a blinking state in such measurements.
To discard these blinking events, we perform an anti-correlation check whenever a state is classified as `$D$' by applying a microwave $\pi$ pulse and verifying that the subsequent readout yields `$B$'.
Note that if the first readout yields `$B$', the \gev center is likely in the correct charge state and this check can therefore be skipped.
The photon detector's dark counts introduce a measured average of $<0.003$ photons per readout and are neglected.
On average, $83.5\,\%$ of `$B$' and $79.5\,\%$ of `$D$' input states pass the anti-correlation check.
To exclude potential bias from applying the anti-correlation check only to `$D$' outcomes, we confirm that applying the check to both `$B$' and `$D$' outcomes yields a comparable SSR fidelity of $96.10\,\%$.
The use of composite pulses and the anti-correlation check reduces the respective SSR infidelities to $P(D|B)=5.65\,\%$ and $P(B|D)=2.74\,\%$, thereby increasing the average SSR fidelity to $\mathcal{F}_{\text{SSR}}^{\,e}=95.80\,\%$. 
Our SSR fidelity is a significant improvement over the previously-reported $74\,\%$ for the \gev center\,\cite{chen2022quantum}. 
Aside from the composite pulse and anti-correlation check, the improved performance likely stems from better magnetic field alignment which maximizes cyclicity, and the superior optical stability of our \gev center.
The SSR fidelity is now mainly limited by the low collection efficiency of the system.
We further calculate the QND fidelity $\mathcal{F}_{\text{QND}}^{\,e}$ of the SSR to be $75.71\,\%$, via~\cite{lupacscu2007quantum}
\begin{equation}
    \mathcal{F}_{\text{QND}} = \frac{P(B^{\text{2nd}}|B^{\text{1st}})}{2} + \frac{P(D^{\text{2nd}}|D^{\text{1st}})}{2},
    \label{eq:qnd_fidelity}
\end{equation}
where the conditional probabilities $P(i|j)$ refer to the probability of the second readout (superscript 2nd) yielding $i$ if the first readout (superscript 1st) yielded $j$.
The $\mathcal{F}_{\text{QND}}^{\,e}$ can be optimized by choosing a shorter readout window of $40\,\text{\textmu s}$ which yields $\mathcal{F}_{\text{QND}}^{\,e} = 90.22\,\%$ with a similarly high $\mathcal{F}_{\text{SSR}}^{\,e} = 89.77\,\%$. 
%

\section{Correlation Spectroscopy and Distant Nuclear Spin Control}

\begin{figure*}
    \centering
    \includegraphics{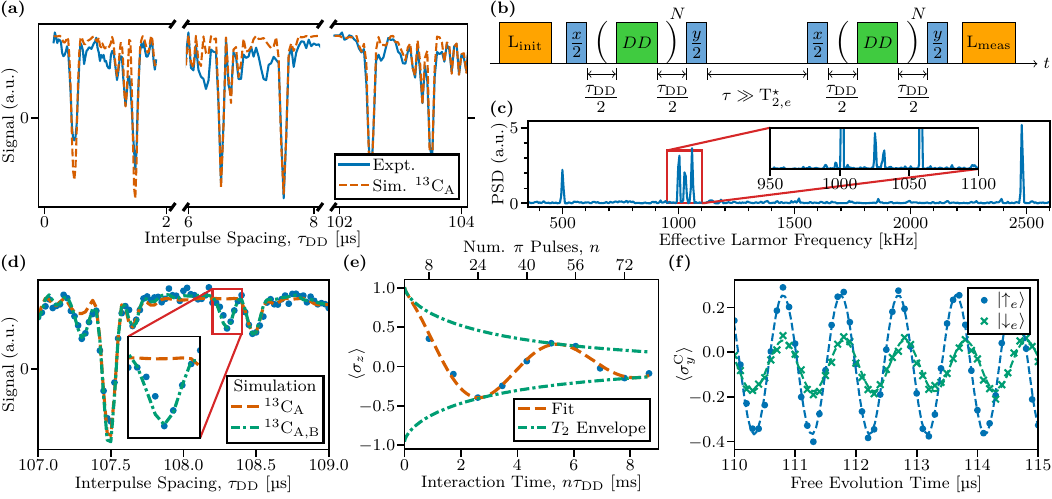}
        \caption{Detection and control of \nuc nuclear spins via dynamical decoupling (DD) and correlation spectroscopy (CS).
        \quad(a) Measured XY8-1 spectrum (blue) at increasingly larger interpulse spacings $\tau_\text{DD}$, probing the \gev center's nuclear spin environment.
        The dotted orange line shows the simulated spectrum which includes the interaction with the strongly coupled nuclear spin ${}^{13}\text{C}_\text{A}$, capturing most features.
        (b) CS pulse sequence: blue blocks indicate microwave $\pi/2$ pulses on the \gev center's $\nu_0$ transition, orange blocks correspond to laser pulses and the green DD block is an XY8 sequence with fixed \taudd and fixed order $N$.
        The sequence induces two conditional rotations of the \gev center's spin state separated by a correlation time $\text{\Ttwostar[e]}\ll\tau<T_{1,e}$ which allows for the detection of the nuclear spin's effective Larmor frequencies.
        %
        %
        %
        (c) Power spectral density (PSD) of the CS measurement, revealing the electron spin state-dependent Larmor frequencies of surrounding \nuc nuclear spins.
        The inset is a zoom-in of a measurement at an optimized \taudd to better resolve the weakly-coupled spins.
        (d) XY8-1 spectrum (blue data points) compared with the simulated spectrum when only considering ${}^{13}\text{C}_\text{A}$ (orange dashed line) and when additionally including ${}^{13}\text{C}_\text{B}$ (green dash-dotted line). 
        The inset showcases a resonance attributed to ${}^{13}\text{C}_\text{B}$.
        (e) XY8-$N$ order sweep for fixed \taudd to probe the coherent interaction with ${}^{13}\text{C}_\text{B}$, which manifests as oscillations in the signal as the number of applied $\pi$ pulses $n$ is increased.
        Additionally, decoherence of the \gev center's electron spin leads to a decay of the signal.
        (f) Ramsey measurement of ${}^{13}\text{C}_\text{B}$ after its measurement-based initialization. 
        The hyperfine interaction with the \gev center results in two distinct Larmor frequencies, $f_L^{\ket{\uparrow_e}}$ for $\ket{\uparrow_e}$ (blue) and $f_L^{\ket{\downarrow_e}}$ for $\ket{\downarrow_e}$ (green).
        }
    \label{fig:fig2}
\end{figure*}

Building a quantum register often requires the precise identification and control of individual nuclear spins within a spin bath.
%
%
Identifying more distant nuclear spins with coupling strengths below $1/$\Ttwostar[e] requires more advanced sensing schemes.
More distant spins are commonly-observed via standard dynamical decoupling (DD) sequences such as XY8~\cite{gullion1990new} but offer only second-order sensitivity to hyperfine interactions~\cite{zahedian24blueprint,beukers2025control,nguyen2019integrated}, resulting in overlapping resonances at short pulse spacings with limited resolvability of individual \nuc nuclear spins.
At longer pulse spacings, this overlap can still occur and the distinguishability of nuclear spins is eventually limited by the electron spin coherence time \Ttwo{e}.
While increasing the number of $\pi$ pulses can enhance the DD spectral resolution, the associated increased heating of the system restricts both the number of pulses that can be applied and their spacing. 
Moreover, strongly-coupled nuclear spins can have additional harmonics which dominate the DD spectrum, masking the \gev center's interactions with more distant spins and making their identification challenging~\cite{casanova2015robust,zahedian24blueprint}.
A typical XY8-$N$ spectrum of our system (XY8 sequence repeated $N$ times, denoted as its order) for varying pulse spacings at fixed order $N=1$ is shown in \autoref{fig:fig2}\,(a).
Most features in the DD spectrum can be attributed to the strongly-coupled nuclear spin (\nucc$_\text{A}$ in \autoref{tab:corrspec_peaks}). \\
These limitations of DD motivate the use of alternative approaches such as correlation spectroscopy (CS) which has been successfully employed to distinguish the contributions of individual nuclear spins in $S=1$ systems in diamond~\cite{laraoui2013high, Staudacher2015natcommun,vorobyov2022addressing,Kong2015prapplied,staudenmaier2023optimal,Degen2017Quantum} and has been proposed as a technique for $S=1/2$ systems as it offers a spectral resolution ultimately limited by \Tone\cite{zahedian24blueprint} which in our system is $20.7\,\text{s}$~\cite{grimm2025coherent}.
The corresponding pulse sequence is shown in \autoref{fig:fig2}\,(b).
After initialization into the dark state, the \gev center is brought to the equatorial plane with a $\pi_x/2$ pulse (index denotes the phase), followed by a DD sequence, which in our case is the XY8-1 sequence.
Together with the subsequent $\pi_y/2$ pulse, this gate conditionally rotates the \gev center if the $\pi$ pulse spacing of the DD block \taudd matches the resonance of one or multiple nuclear spins~\cite{taminiau2012detection,zahedian24blueprint}.
Repeating this gate sequence after a correlation time $\tau$ with $\text{\Ttwostar[e]}\ll\tau<T_{1,e}$~\cite{zahedian24blueprint} allows for the correlation of the conditionally rotated states of the \gev center which can then be read out via a laser pulse.
%
%
By varying $\tau$ and taking the Fourier transform of the signal the individual nuclear precession frequencies $f_L^{\ket{\uparrow_e}}$ and $f_L^{\ket{\downarrow_e}}$ can be observed~\cite{laraoui2013high, Staudacher2015natcommun, Kong2015prapplied,vorobyov2022addressing,staudenmaier2023optimal,Degen2017Quantum}. \\
Numerical simulations indicate that residual nuclear spin coherence across consecutive measurement steps can produce additional harmonics in the Fourier transform~\cite{supplement}.
While initializing the nuclear spins prior to each cycle can resolve this problem, it would introduce significant experimental overhead and complexity.
Instead, simulations show that randomizing the order in which $\tau$ is swept during the measurement is sufficient to convert spurious correlations between successive measurements into uncorrelated noise when the signal is reordered prior to the Fourier transform~\cite{supplement}. \\
Experimentally, we use Welch's method to estimate the power spectral density (PSD) of the measured signal as it provides superior noise robustness compared to a typical fast Fourier transform~\cite{welch2003use, harris2005use, stoica2005spectral}.
We perform CS at an XY8 $\pi$ pulse spacing of $\text{\taudd}=10.19\,\text{\textmu s}$ (see \autoref{fig:fig2}\,b) to obtain an initial signal from multiple nuclear spins and then fine-tune \taudd to $10.22\,\text{\textmu s}$ to better resolve distant spins.
The result is shown in \autoref{fig:fig2}\,(c), where the outermost two peaks correspond to the effective Larmor frequencies of the strongly-coupled \nuc nuclear spin (\nucc$_\text{A}$ in \autoref{tab:corrspec_peaks})
and the remaining four peaks (inset of \autoref{fig:fig2}\,(c)) correspond to two weakly-coupled \nuc nuclear spins.
Note that while the hyperfine interaction of the \gev center with \nuc nuclear spins can be described by the hyperfine vector $\mathbf{A}=    (A_{zx},A_{zy},A_{zz})^\top$~\cite{casanova2015robust}, we report the values in rotated coordinate systems where each $A_{zy}=0$.
In order to obtain these results, we extend the CS method to a two-dimensional measurement where both \taudd and $\tau$ are varied (see Methods).
CS allows for the immediate extraction of the resonance frequencies of the individual nuclear spins (see \autoref{tab:corrspec_peaks}) which can then be used to directly control them. \\

\begin{table}[t]
    \centering
    \caption{
    Nearby \nuc nuclear spins with their distinct Larmor frequencies ($f_L^{\ket{\uparrow_e}}$ and $f_L^{\ket{\downarrow_e}}$) observed in correlation spectroscopy and their hyperfine coupling values $A_{zx}, A_{zz}$ obtained by fitting the XY8 spectrum for a magnetic field of $96.837\,\text{mT}$ as shown in \autoref{fig:fig2}\,(d).
    The contribution from \nucc$_\text{C}$ cannot be distinguished in these measurements and we therefore omit its hyperfine coupling estimation.
    }
    \begin{tabular}{ccccc}
        \toprule
        & $f_L^{\ket{\uparrow_e}}$ (kHz) & $f_L^{\ket{\downarrow_e}}$ (kHz) & $A_{zx}/2\pi$ (kHz) & $A_{zz}/2\pi$ (kHz) \\
        \hline
        \nucc$_\text{A}$ & 499.13 & 2479.38 & 598 & -2963 \\
        \nucc$_\text{B}$ & 1001.65 & 1058.62 & 128 & 39 \\
        \nucc$_\text{C}$ & 1025.39 & 1032.17 & - & - \\
        \hline
    \end{tabular}
    \label{tab:corrspec_peaks}
\end{table}
%
%
This method also enables confident attribution of features in the DD spectrum to individual \nuc nuclear spins by verifying that the corresponding CS spectrum exhibits only two frequencies.
For example in \autoref{fig:fig2}\,(d), the orange dashed line shows the simulated XY8-1 spectrum with only the strongly-coupled \nuc nuclear spin (\nucc$_\text{A}$), and the green dash-dotted line shows the result when the second \nuc nuclear spin (\nucc$_\text{B}$) is also considered. 
The hyperfine coupling values are obtained by fitting simulations to the measured XY8 spectrum and are reported in \autoref{tab:corrspec_peaks}. \\ 
Increasing the order $N$ of XY8 on the \taudd which is selective to \nucc$_\text{B}$ leads to an oscillation of the \gev center's spin state as shown in \autoref{fig:fig2}\,(e). 
This oscillation corresponds to the rotation induced on the nuclear spin if the \gev center is continuously flipped between $\ket{\uparrow_e}$ and $\ket{\downarrow_e}$ and can be used to indirectly drive nuclear Rabi oscillations\,~\cite{taminiau2012detection,supplement}.
Maximum entanglement between the \gev center and \nucc$_\text{B}$ can then be achieved by choosing the order corresponding to the zero-crossing of $\langle{\sigma_z}\rangle$ in \autoref{fig:fig2}\,(e).
Sandwiching the appropriate XY8-$N$ gate between a $\pi_x/2$ and $\pi_y/2$ pulse (as in CS) allows to realize a basis-transformed $\text{C}_{n}\text{NOT}_e$ gate~\cite{nguyen2019integrated, taminiau2014universal, klotz2025bipartite}, which flips the \gev center if the \nuc nuclear spin is in $\ket{\text{-}y} = (\ket{\uparrow_n} - i \ket{\downarrow_n}) / \sqrt{2}$ and leaves it unchanged if the \nuc nuclear spin is $\ket{y}$. 
Combining this gate with the SSR introduced in the previous section enables measurement-based initialization (MBI) of the nuclear spin.
Assuming the \nuc nuclear spin is initially in a thermal equilibrium $\rho_n=(\ket{y}\bra{y}+\ket{\text{-}y}\bra{ \text{-}y})/2$, an SSR outcome yielding `$D$' (i.e. $\ket{\downarrow_e}$) projects the \nuc nuclear spin state into $\ket{y}$, while `$B$' (i.e. $\ket{\uparrow_e}$) projects it into $\ket{\text{-}y}$. \\
MBI allows us to directly prepare \nucc$_\text{B}$ in a superposition state and perform a Ramsey measurement by simply varying the free evolution time before performing a subsequent $\text{C}_{n}\text{NOT}_e$ gate and a readout laser pulse.
The measurement result in \autoref{fig:fig2}\,(f) shows the nuclear spin precessing at either $f_L^{\ket{\uparrow_e}}$ or $f_L^{\ket{\downarrow_e}}$ depending on the electron spin state. 
We obtain a \Ttwostar[n] of $(1.98\,\pm\,0.15)\,\text{ms}$ (see Methods).
As we operate at a high XY8 pulse spacing of $\tau_\text{DD}=108.29\,\text{\textmu s}$, the signal's amplitude is severely decreased due to the electron spin's decoherence during the $\text{C}_{n}\text{NOT}_e$ duration of $\approx1.8\,\text{ms}$ which is close to the \Ttwo{e} of $(3.52\,\pm\,0.35)\,\text{ms}$.
While resonances at much lower $\tau_\text{DD}$ could therefore enhance the contrast, the competing effect of higher heating makes this challenging to observe experimentally.
We also note that the measured signals are offset relative to the nuclear spin expectation value $\langle{\sigma_y^\text{\nucc}}\rangle = 0$ and have different amplitudes.
Numerical simulations (see Methods) show that small shifts of the XY8 resonances over the course of the experiment can cause such behavior and we attribute these shifts to magnetic field instabilities~\cite{grimm2025coherent}.

\section{Single-shot Readout on a \nuc nuclear spin}

\begin{figure}
    \centering
    \includegraphics{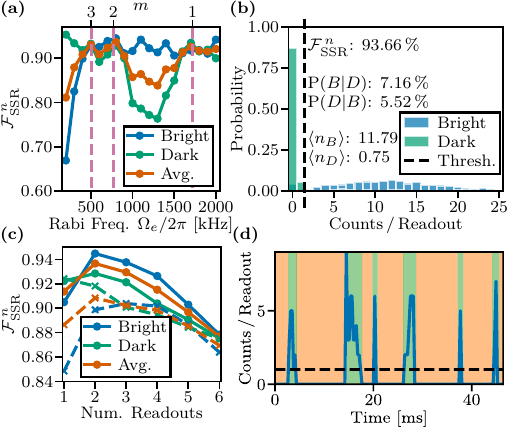}
        \caption{
    Optimizing single-shot readout on the strongly-coupled ${}^{13}\text{C}_\text{A}$ nuclear spin.
    \quad(a) Nuclear SSR fidelity for different Rabi frequencies $\Omega_e$ of the $\text{C}_{n}\text{NOT}_e$ gate. 
    The pink dashed line marks frequencies (with index $m$) at which the $\text{C}_{n}\text{NOT}_e$ gate (i.e. a $\pi$ pulse on the \gev center at $\nu_1$ or $\nu_2$), induces an $m \cdot 2\pi$ rotation on the corresponding other hyperfine transition.  
    The SSR of the bright \nuc spin state is shown in blue, the dark \nuc spin state in green and the average fidelity in orange.
    (b)	Nuclear SSR histogram for $m=2$, yielding $\mathcal{F}_{\text{SSR}}^{\,n}=93.66\,\%$.
    The readout is repeated twice with an optimal photon threshold (black dashed line) of two.
    The resulting dark state ($\ket{\downarrow_e}$) photon counts are shown in green, and bright state ($\ket{\uparrow_e}$) in blue.
    (c) Optimization of $\mathcal{F}_{\text{SSR}}^{\,n}$ by repetitive nuclear spin readout.
    The fidelity decreases for repetitions $>2$ due to back action of the $\text{C}_{n}\text{NOT}_e$ gate on the nuclear spin state.
    The solid lines correspond to the optimized setting at $m=2$, while the dashed lines show the fidelity when operating at a Rabi frequency of $\Omega_e/(2\pi) = 400\,\text{kHz}$, slightly above the \Ttwostar[e] limit.
    (d) Observation of quantum jumps on the nuclear spin during the repetitive readout in (c).
    }
    \label{fig:fig3}
\end{figure}

The high $\mathcal{F}^{\,e}_{\text{SSR}}$ on the \gev center further permits SSR on the strongly-coupled \nuc nuclear spin.
Here, the readout consists of a $\text{C}_{n}\text{NOT}_e$ gate implemented by a weak, narrow-band microwave $\pi$ pulse at a nuclear spin hyperfine transition (i.e. $\nu_1$, $\nu_2$ in \autoref{fig:fig1}\,(a)), followed by a readout laser pulse.
While a lower Rabi frequency reduces power broadening and can therefore increase the selectivity of the microwave pulse, it increases the susceptibility to low frequency noise. 
This sets a practical lower bound for the microwave pulse length, given by \Ttwostar[e]~\cite{senkalla2024germanium}, at $\sim(2\pi)\,350\,\text{kHz}$.
Additionally, we observe spectral diffusion of the $\text{C}_{n}\text{NOT}_e$ transitions of $\pm\,150\,\text{kHz}$, resulting in significant detuning that reduces the nuclear $\mathcal{F}_{\text{SSR}}^{\,n}$. \\
Although larger Rabi frequencies can partly compensate these effects, the associated power broadening results in an increased overlap with the other hyperfine transition.
It is possible, however, to analytically derive higher Rabi frequencies that induce a detuned $2\pi$ rotation on the off-resonant transition and ensure selectivity of the $\pi$ pulse~\cite{supplement}.
Such a Rabi frequency is given by $\Omega_{e,2\pi}(m) = {\Delta} / {\sqrt{4m^2-1}}$ with $m \in \mathbb{Z^+}$ and the detuning $\Delta\mathrel{\hat=}\vert{A_{zz}}\vert=(2\pi)\,2963\,\text{kHz}$.
To verify this experimentally, we sweep the Rabi frequency of the readout $\text{C}_{n}\text{NOT}_e$ gate and plot the resulting $\mathcal{F}_{\text{SSR}}^{\,n}$ in \autoref{fig:fig3}\,(a).
Note that the fidelity is calculated using an optimized total photon threshold of two using two consecutive readouts~\cite{supplement}.
The \nuc nuclear spin is initialized via our previously-reported spin pumping scheme~\cite{grimm2025coherent}, exploiting the strong $A_{zx}$ interaction.
In general, the bright state fidelity is largely insensitive to $\Omega_e$, as higher Rabi frequencies merely cause unconditional spin flips of the \gev center.
Since the \gev center is initialized in the dark state, a strong $\pi$ pulse generally yields a `$B$' outcome.
However, at low $\Omega_e$, decoherence dominates in our system, significantly reducing the $\text{C}_{n}\text{NOT}_e$ fidelity.
In contrast, the `$D$' state fidelity increases at low $\Omega_e$ since the detuned low-fidelity $\text{C}_{n}\text{NOT}_e$ fails to induce a spin flip (conditional or unconditional) on the \gev center, and the measurement approaches the \gev center’s native initialization fidelity.
We therefore focus on the average state fidelity $\mathcal{F}_{\text{SSR}}^{\,n}$, which is maximized near a detuned $2\pi$ Rabi oscillation (denoted by the pink vertical dashed lines).
Using the optimized Rabi frequency of $\Omega_{e,2\pi}=(2\pi)\,800\,\text{kHz}$, we achieve a maximum nuclear $\mathcal{F}_{\text{SSR}}^{\,n}$ of $93.66\,\%$ see \autoref{fig:fig3}\,(b), approaching the limit set by the \gev center and nuclear spin initialization fidelities of $98\,\%$ and $95\,\%$ respectively ~\cite{grimm2025coherent}.
This fidelity is achieved without the use of the anti-correlation check on the \gev center discussed earlier.
We note that the optimized $\Omega_{e,2\pi}$ deviates slightly from the analytical value of $\Omega_{e,2\pi}(m=2) \approx (2\pi)\,765\,\text{kHz}$ due to the finite resolution of the microwave source as well as uncertainty in the hyperfine parameters of ${}^{13}\text{C}_\text{A}$. \\
The variation in fidelities with consecutive readouts are shown in \autoref{fig:fig3}\,(c), with the results for the optimal Rabi frequency $\Omega_{e,2\pi}$ shown as solid lines and a comparison with the lower $\Omega_e = (2\pi)\,400\,\text{kHz}$ (dashed lines, peak average fidelity $90.84\,\%$) near the \Ttwostar[e] limit, visualizing the clear benefit of the optimized $\Omega_{e,2\pi}$.
After two readouts, the fidelity drops due to the strong $A_{zx}$ interaction which causes the $\text{C}_{n}\text{NOT}_e$ gate to have significant back action on the nuclear spin~\cite{supplement}.
%
%
For a single readout of the nuclear spin, we achieve an $\mathcal{F}_{\text{SSR}}^{\,n}$ of $90.59\,\%$ 
and a corresponding $\mathcal{F}_{\text{QND}}^{\,n}$ of $80.52\,\%$.
The high $\mathcal{F}_{\text{SSR}}^{\,n}$ enables the observation of quantum jumps of the nuclear spin  (e.g. due to the $\text{C}_{n}\text{NOT}_e$ gate back action) as shown in \autoref{fig:fig3}\,(d).
%

\section{Conclusion and outlook} 

We demonstrate single-shot readout (SSR) of a single \gev center in diamond with up to $95.80\,\%$ fidelity.
This significantly surpasses the previously-reported fidelity of $74\,\%$~\cite{chen2022quantum} and is facilitated by a robust composite pulse that mitigates state preparation errors, an anti-correlation check that rejects blinking events, and an optimally-aligned magnetic field. 
Such high-fidelity SSR is crucial for error detection, correction and feed-forward operations in quantum networks~\cite{abobeih2022fault,salathe2018low,xiang2020simultaneous}. \\
We further show that correlation spectroscopy (CS) can identify multiple nuclear spins in a convoluted noise with overlapping resonances in the \gev center's dynamical decoupling (DD) spectrum, thus overcoming a major drawback of spin-$1/2$ systems.
DD allows for the control of the nuclear spins and creation of a basis-transformed $\text{C}_{n}\text{NOT}_e$ gate~\cite{nguyen2019integrated, taminiau2014universal, klotz2025bipartite} which, combined with the \gev center's high fidelity single-shot readout, facilitates the measurement-based initialization of the nuclear spins and observation of their effective Larmor frequencies.
The method of identifying, controlling and initializing distant \nuc nuclear spins can be extended to dozens of spins~\cite{bradley2019ten,van2024mapping}, allowing to efficiently scale the spin register. 
Such register are crucial to preserve information during entanglement with other network nodes~\cite{knaut2024entanglement, briegel1998quantum, kalb2017entanglement} and are required for error correction schemes~\cite{chang2025hybrid, unden2016quantum, chao2018quantum}. 
Combining radio-frequency control with DD gates~\cite{beukers2025control} can further improve gate fidelities and enhance signal strength in CS. \\
Additionally, we demonstrate SSR of a nearby \nuc nuclear spin by combining the \gev center's SSR with a $\text{C}_{n}\text{NOT}_e$ gate implemented through a narrow-band microwave $\pi$-pulse.
By tuning the microwave amplitude to induce a $\pi$-rotation on a single hyperfine transition while simultaneously applying a $2\pi$-rotation on the other, we achieve a SSR fidelity of $93.66\,\%$, close to the estimated limit set by the initialization fidelities of the \gev center and nuclear spin.
The $\mathcal{F}_{\text{SSR}}^{\,n}$ and $\mathcal{F}_{\text{QND}}^{\,n}$ fidelities can be further enhanced by optimizing the $\text{C}_{n}\text{NOT}_e$ gate through optimal control~\cite{muller2022one,rossignolo_quocs_2023,vetter2024gate}. \\
Our work significantly advances the applicability of the \gev center and surrounding nuclear spins towards a fully operational quantum network node. The techniques demonstrated are general and thus directly applicable to other spin-$1/2$ systems.
%

\section{Methods}

\subsection{Hamiltonian of the System}

Due to the high magnetic field, the system's dynamics for $j$ nuclear spins can be described in the rotating frame of the GeV Larmor frequency $\omega_0= \gamma_{{}\text{GeV}} \,B_z$, where $\gamma_{{}\text{GeV}}$ denotes the gyromagnetic ratio of the GeV electron spin and $B_z$ is the magnitude of the magnetic field along the $z$ direction.
Under the secular approximation~\cite{maze2012free} and using the rotating wave approximation, the system's Hamiltonian is given by
\begin{equation}
    \begin{split}
        H = ~&\Delta S_z +\Omega_e \left(S_x \sin{\theta} + S_y \cos{\theta} \right) - \sum_j 
        \omega_L I_z^j\\
        &+ S_z\sum_j \left(A_{zx}^jI_x^j + A_{zy}^jI_y^j + A_{zz}^jI_z^j\right) 
    \end{split}
    \label{eq:hyperfine_interaction}
\end{equation}
with the electron (nuclear) spin operators $S_{x,y,z}~(I_{x,y,z})$, the hyperfine coupling parameters $A_{zx}$, $A_{zy}$ and $A_{zz}$, the Rabi frequency $\Omega_e$ of linearly polarized microwave fields with a phase $\theta$ and detuning $\Delta$ from the \gev center's Larmor frequency $\omega_0$, with the nuclear Larmor frequency $\omega_L=\gamma_{{}^{13}\text{C}} B_z$, where $\gamma_{{}^{13}\text{C}}$ denotes the gyromagnetic ratio of a \nuc nuclear spin. 
Note that for each nuclear spin given in \autoref{tab:corrspec_peaks}, one can choose a rotated coordinate system such that $A_{zy}=0$.

\subsection{GeV Center Blinking and Estimation of Blinking Rates}

\begin{figure}[t]
    \centering
    \includegraphics{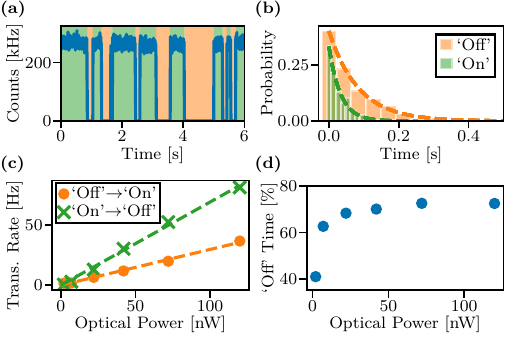}
        \caption{
        Optical power-dependent blinking behavior of the \gev center in the absence of a magnetic field.
        \quad(a) Example of blinking of the \gev center at an optical power of $7\,\text{nW}$.
        The `off' state denotes the regions of low ($\sim0$) photon count rates (shaded orange) and the `on' state denotes regions of high photon count rates (shaded green).
        (b) Example of a histogram of the time spent in the `on' and `off' states at an optical power of $42\,\text{nW}$.
        The histograms are obtained by binning the durations of the `off' and `on' states in (a). 
        A simple exponential decay is fit (dashed lines) to the histograms to extract a characteristic `on' and `off' time for a specified optical power.
        (c) The `on'$\rightarrow$`off' and `off'$\rightarrow$`on' transition rates as a function of optical power, indicating a linear relation with the gradients $m_\text{`on'$\rightarrow$`off'} = (0.701\pm0.022)\,\text{nW\,/\,Hz}$ and $m_\text{`off'$\rightarrow$`on'} = (0.297\pm0.011)\,\text{nW\,/\,Hz}$.
        The `on'$\rightarrow$`off' (`off'$\rightarrow$`on') transition rate is defined as the inverse of the characteristic `on' (`off') time determined in (c).
        (d) The average percentage of time the \gev center spends in the ‘off’ state at different optical powers. 
        For all other measurements presented in this work, the optical powers used were below the lowest value here of $2\,\text{nW}$.
        }
    \label{fig:fig4}
\end{figure}
Empirically, we have observed the \gev center in this work recovering from blinking events under resonant laser illumination and without the need for an additional off-resonant re-pump laser.
We have also observed the rate of the onset and recovery from blinking events varying with input optical power.
We therefore study the optical power dependence of the blinking rates without a magnetic field at 4\,K to allow clear distinguishing of blinking events from spin-pumping initialization of the \gev center.
We define the `on' state of the \gev center as the optically active state that we use for experiments and the `off' state as the blinking state where no fluorescence is detected from the \gev center. 
This allows the definition of two transition rates; the `on'$\rightarrow$`off' rate which describes the transition rate between the optically active state of the \gev center to the blinking state and the `off'$\rightarrow$`on' state which describes the converse rate. \\
To determine these rates, we measure the fluorescence of the \gev under resonant excitation at varying optical powers and bin the times taken to transition to and from the `on' and `off' states.
In \autoref{fig:fig4}\,(a), an example of the count rate trace used to determine the transition rates at an optical power of $7\,\text{nW}$ is shown.
The `off' state denotes the regions of low ($\sim0$) photon count rates (shaded orange) and the `on' state denotes regions of high photon count rates (shaded green).
By binning these `on' and `off' times, a histogram is obtained as shown in \autoref{fig:fig4}\,(b).
Fitting a simple exponential decay to these histograms allows the transition rates at given optical powers to be determined. \\
Repeating this measurement for various optical powers enables the dependence of the transition rates on laser power to be determined as shown in \autoref{fig:fig4}\,(c).
The lowest optical power in the measurements shown is $2\,\text{nW}$ and the highest is $120\,\text{nW}$.
%
A linear fit of the form $Y=m\cdot X$ yields the gradients $m_\text{`on'$\rightarrow$`off'} = (0.690\pm0.014)\,\text{nW\,/\,Hz}$ and $m_\text{`off'$\rightarrow$`on'} = (0.2966\pm0.0066)\,\text{nW\,/\,Hz}$.
The strong linear correlation ($R^2 > 0.99$ for both fits) suggests that the processes driving the \gev center into and out of the `off' state are most likely single-photon processes.
This is in contrast to what \autoref{fig:fig4}\,(d) suggests, which is a non-linear dependence of the average `off' time percentage on optical power.
If both transitions were single-photon processes, we would expect the total `off' time percentage to be invariant to optical power. \\
For a multi-photon process, it is generally expected that each individual photon transition will have its own saturation power, causing the low- and high-optical power regimes to have different trends, e.g, initially quadratic/ exponential before  becoming linear when all but one process have been saturated.
We do not exclude the possibility that we are already saturating multi-photon transitions even at the lowest power used and are therefore already in/ near the linear regime.
Measurements at powers below $1\,\text{nW}$ are needed to verify this hypothesis but current experimental limitations such as low count rates and optical power fluctuations prevent us from accurately conducting these measurements.
The distinct gradients also suggest that as optical power increases, the \gev center is likely to spend more time in the `off' state, as illustrated in \autoref{fig:fig4}\,(d), which shows the `off' time versus the overall measurement time. 
\subsection{Two-Dimensional Correlation Spectroscopy}

\begin{figure}[b]
    \centering
    \includegraphics{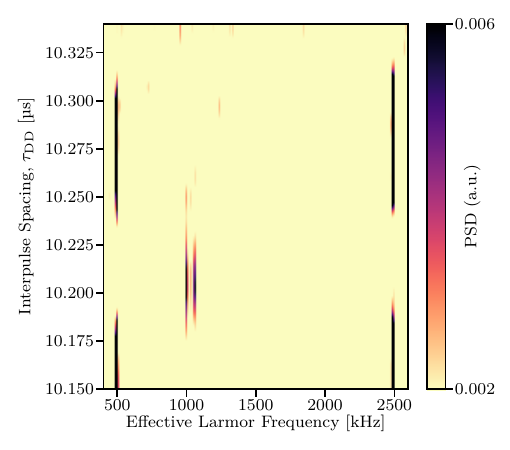}
        \caption{
        Two-dimensional correlation spectroscopy where $\tau$ is swept for 20 consecutive \taudd values.
        The color scale indicates the power spectral density (PSD) obtained via Welch's method.
        %
        }
    \label{fig:fig5}
\end{figure}

\begin{figure}
    \centering
    \includegraphics{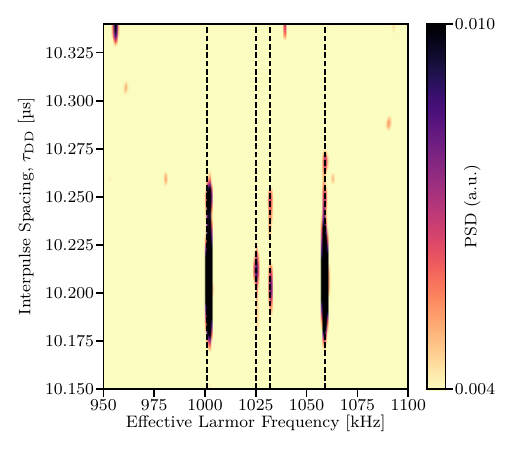}
        \caption{
        Two-dimensional correlation spectroscopy measurement in \autoref{fig:fig5} but with the plotted frequency range reduced to better visualize the weakly-coupled \nuc spins.
        The dashed lines indicate the positions of \nucc$_\text{B}$ wand \nucc$_\text{C}$.
        }
    \label{fig:fig5_zoom}
\end{figure}
The method of correlation spectroscopy (CS) \cite{laraoui2013high,Kong2015prapplied,Staudacher2015natcommun} traditionally includes variation of the time separation $\tau$ between the two DD sequences (see Fig. \ref{fig:fig2}).
We extend the method to two dimensions and scan both $\tau$ and the inter-pulse separation $\tau_{\text{DD}}$.
Such a two-dimensional correlation spectroscopy measurement is shown in \autoref{fig:fig5} where 4096 $\tau$ values are measured for 20 consecutive \taudd values, with a $\tau$ step size of $180\,\text{ns}$ and \taudd step size of $10\,\text{ns}$.
We note that the well-defined measurement results are indicating that shot noise is not significant with a large number of $\tau$ steps, despite only having performed 30 measurement shots for each of the $4096 \times 20 = 81920$ measured $\tau$ points.
This strategy of using a large number of $\tau$ steps to mitigate shot noise has the added benefit of enabling a high frequency resolution since the number of $\tau$ steps is what defines the frequency resolution of the Fourier transform, with the $\tau$ step size determining the highest frequency resolvable.
This further suggests that such a two-dimensional correlation spectroscopy measurement is a time-efficient way to identify \taudd that are selective to single \nuc spins, especially in the presence of a strongly-coupled \nuc spin.
To better resolve the weakly-coupled spins, a plot of the same measured data over a limited frequency range is shown in \autoref{fig:fig5_zoom}.
Note that since the high frequency resolution of the measurement results in very fine and narrow vertical lines, a Gaussian interpolation is used to enhance their visibility.
\subsection{Nuclear Ramsey Simulations}

\begin{figure}[t]
    \centering
    \includegraphics{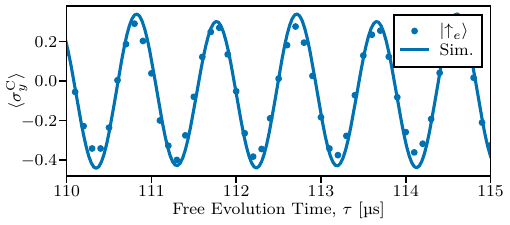}
        \caption{Nuclear Ramsey simulation.
        Simulating the nuclear Ramsey measurement sequence (solid line) including the measurement-based initialization using a \taudd that is shifted approx $35\,\text{ns}$ from the resonance can very closely replicate the measured signal for the $\ket{\uparrow_e}$ case.}
    \label{fig:fig6}
\end{figure}
To investigate the offset of the nuclear Ramsey signal amplitude relative to the nuclear spin expectation value $\langle{\sigma_y^\text{\nucc}}\rangle = 0$ as seen in \autoref{fig:fig2}\,(f), we simulate the nuclear Ramsey measurement sequence including the measurement-based initialization.
We find that using a \taudd that is shifted $\approx 35\,\text{ns}$ from the optimal resonance dip identified via XY8 can replicate the measured signal for the $\ket{\uparrow_e}$ case as shown in \autoref{fig:fig6}.
The exact cause of these shifts requires further investigation but are likely due to instability of the superconducting vector magnet~\cite{grimm2025coherent}.
The use of a high \taudd exacerbates the effects of such instabilities as the selectivity of the DD sequence increases with higher \taudd \cite{Degen2017Quantum}, resulting in an increased sensitivity to such spectral shifts.  
In the $\ket{\uparrow_e}$ manifold, we observe a similar behavior but with the amplitude further decreased.
We attribute this effect to the poorer single-shot readout fidelity for the $\ket{\uparrow_e}$ state (i.e. the `D' state) as blinking events can be misconstrued for the `D' state as discussed in earlier sections.
\begin{figure}
    \centering
    \includegraphics{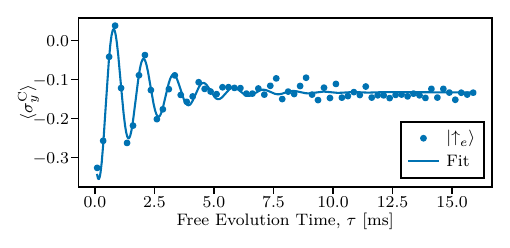}
        \caption{Undersampled nuclear Ramsey measurement on \nucc$_\text{B}$, yielding a \Ttwostar[n] of $(1.98\,\pm\,0.15)\,\text{ms}$.}
    \label{fig:fig7}
\end{figure}
We measure the \Ttwostar[n] of \nucc$_\text{B}$ using an undersampled Ramsey measurement and obtain a \Ttwostar[n] of $(1.98\,\pm\,0.15)\,\text{ms}$ which is comparable to the previously-reported  \Ttwostar[n] of \nucc$_\text{A}$ of $(2.00\,\pm\,0.18)\,\text{ms}$\cite{grimm2025coherent}.

\begin{acknowledgments}
    We thank Vadim Vorobyov, Lev Kazak, Marco Klotz, Andreas Tangemann, Thomas Reisser and Jurek Frey for helpful discussions.
    We also thank Yuri N. Palyanov, Igor N. Kupriyanov and Yuri M. Borzdov for providing the sample used in this work. 
    The authors acknowledge support by the state of Baden-W\"urttemberg through bwHPC.\\
    We acknowledge BMFTR for support via projects SPINNING (No. 13N16215), DE BRILL (No. 13N16207), Quanten4KMU (No. 03ZU1110BB), QuantumHiFi (No. 16KIS1593), QR.N, CoGeQ and Deutsche Forschungsgemeinschaft (DFG) via project No. 387073854 and joint DFG-Japan Science and Technology Agency (JST) project ASPIRE (No. 554644981).
    
\end{acknowledgments}


\section{Author Contributions}
\label{sec:author_contributions}

P.G., K.S., and P.J.V. conceived and planned the experiments.
K.S. initially built the experimental setup, with later contributions from P.G..
P.G performed the measurements, data analysis and simulations.
K.S. additionally assisted in finding isolated resonances in the dynamical decoupling spectrum.
P.G., P.J.V., and K.S. jointly interpreted the results.
M.O.-F. and P.W. contributed to the interpretation of the blinking behavior.
G.G. and M.M.M. supported the simulations and discussions.
P.J.V., P.G., and K.S. wrote the main manuscript with input from all authors.
All work was supervised by K.S., P.J.V. and F.J.
All authors discussed the findings and approved the final version of the manuscript.

\section{Data Availability}
\label{sec:data_availability}

The data presented in this study is available from the corresponding authors on reasonable request.

\newpage
\bibliographystyle{ieeetr}
\bibliography{references}


\end{document}



\title{Supplementary Information: High-Fidelity Single-Shot Readout and Selective Nuclear Spin Control for a Spin-1/2 Quantum Register in Diamond}

\author{Prithvi Gundlapalli}
\email{prithvi.gundlapalli@alumni.uni-ulm.de}
\affiliation{Institute for Quantum Optics, Ulm University, Albert-Einstein-Allee 11, D-89081 Ulm, Germany}
\author{Philipp J. Vetter}
\affiliation{Institute for Quantum Optics, Ulm University, Albert-Einstein-Allee 11, D-89081 Ulm, Germany}
\author{Genko Genov}
\affiliation{Institute for Quantum Optics, Ulm University, Albert-Einstein-Allee 11, D-89081 Ulm, Germany}
\author{Michael Olney-Fraser}
\affiliation{Institute for Quantum Optics, Ulm University, Albert-Einstein-Allee 11, D-89081 Ulm, Germany}
\author{Peng Wang}
\affiliation{Institute for Quantum Optics, Ulm University, Albert-Einstein-Allee 11, D-89081 Ulm, Germany} 
\author{Matthias M. M\"uller}
\affiliation{Peter Gr\"unberg Institute-Quantum Control (PGI-8), Forschungszentrum J\"ulich GmbH, D-52425 J\"ulich, Germany}
\author{Katharina Senkalla}
\affiliation{Institute for Quantum Optics, Ulm University, Albert-Einstein-Allee 11, D-89081 Ulm, Germany}
\author{Fedor Jelezko}
\email{fedor.jelezko@uni-ulm.de}
\affiliation{Institute for Quantum Optics, Ulm University, Albert-Einstein-Allee 11, D-89081 Ulm, Germany}
\date{\today}
    
\maketitle

\tableofcontents

\section{Experimental Setup}\label{si:expt_setup}
The sample is in a dilution refrigerator (Bluefors BFLD400) with a home-built $4\text{f}$ confocal microscope for optical excitation and detection.
A $xyz$ positioner system (Attocube ANPx101/z102/RES, Titanium) and a fine $z$ piezo stage (Attocube ANPz101std, Copper Beryllium) is used to position the sample \cite{senkalla2024germanium}.
A superconducting vector magnet (American Magnetics Inc.) provides a magnetic field of up to $1\,\text{T}$ in the $x,y$ plane and up to $3\,\text{T}$ in the $z$ plane (in the lab frame), enabling the magnetic field to be rotated arbitrarily.
While the magnet can be operated in persistent mode, it was observed to have a decay of approximately $4.7\,\text{mT/h}$~\cite{grimm2025coherent}.
Therefore the magnet was operated in a closed-loop mode which results in slight fluctuations of the magnetic field of approximately $5\,\text{\textmu T}$~\cite{grimm2025coherent} due to the stability of the power supplies.
For most experiments in this work, tunable solid-state laser (C-WAVE GTR, H\"UBNER Photonics) was used to generate CW laser light
at approximately $602\,\text{nm}$ for resonant excitation of the \gev center.
Some initial experiments were performed with a dye laser (Sirah Matisse 2 DS with Rhodamine 6G in ethylene glycol).
We note that experiments performed with the dye laser and solid-state laser yielded comparable results and we therefore do not distinguish between the experiments performed with either laser.
An acousto-optic modulator (Crystal Technology 3200-146) was used to generate laser pulses.
A cleanup filter (Chroma 599/13 ET Bandpass) was used to reject any autofluorescence from the fiber (Thorlabs PM-S405-XP) delivering the laser light to the setup.
A scanning mirror (PI S-335.2SH) was used to scan across the sample with a beam sampler (Thorlabs BSF-20A) used to sample the excitation beam.
The light is then directed into an objective lens (Newport LIO-60x) with a numerical aperture (NA) of 0.85.
The \gev center's phonon sideband emission is detected using a long-pass filter (Chroma 610 LP ET) to minimize the amount of scattered laser light reaching the avalanche photodiode (APD) (Excelitas SPCM-AQRH-16, rated dark count of $25\,\text{Hz}$).
A time-to-digital converter (Fastcomtec MCS6A-1T2) was used to time-tag the APD clicks with a resolution of $200\,\text{ps}$.
All measurements were orchestrated using an open-source Python-based control software Qudi~\cite{binder2017qudi}.
An arbitrary waveform generator (AWG) (Keysight M8195A) was used to generate microwave (mw) pulses.
The spin transitions can be coherently driven using microwaves delivered through a $20\,\text{\textmu m}$ diameter copper wire spanned across the diamond.
For the strong Rabi drive, the mw pulses are sent through a $3\,-\,3.5\,\text{GHz}$ cavity bandpass filter (Mini-Circuits ZVBP-3R25G-S+) to suppress the AWG's noise floor, followed by a circulator (Fairview Microwave FMCR1004) and then amplified using a $50\,\text{W}$ amplifier (Amplifier Research 50S1G6).
For the weak Rabi drive, the mw pulses are sent directly from the AWG to a $1\,\text{W}$ low-noise amplifier (Mini-Circuits ZX60-63GLN+).
The strong and weak Rabi drive mw paths are combined via a $180\,\si{\degree}$ hybrid coupler (Fairview Microwave FMCP1155).
The sum-output of the hybrid coupler then passes through a $2\,-\,4\,\text{GHz}$ suspended substrate stripline bandpass filter (Mini-Circuits ZBSS-3G-S+) to suppress broadband noise introduced by the amplifiers before being sent into the dilution fridge.
Rectangular pulses are used in all measurements in this experiment.

\section{Cyclicity Measurement}

\begin{figure}
    \centering
    \includegraphics{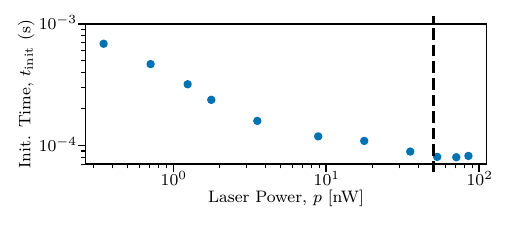}
        \caption{
        Determining $t_{\text{sat}}$ by measuring the mean initialization time at different laser powers $p$.
        The vertical dashed line indicates the saturation laser power regime $p \geq p_\text{sat}$ where $t_{\text{init}}$ plateaus and reaches $t_{\text{sat}}$.
        }
    \label{fig:fig9_init}
\end{figure}

\begin{figure}
    \centering
    \includegraphics{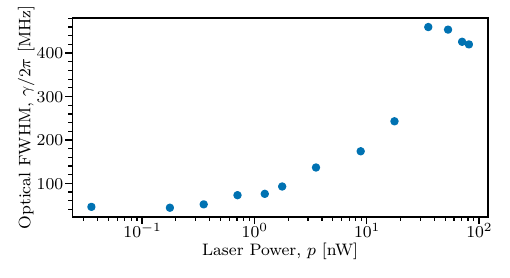}
        \caption{
        Full-width at half maximum (FWHM) of the optical transitions of the \gev center measured at different optical powers $p$.
        At low $p$, the lifetime-limited linewidth of approximately $44\,\text{MHz}$ is obtained.
        }
    \label{fig:fig9_fwhm}
\end{figure}

The cyclicity of our \gev center is estimated by following the procedure described in \cite{rosenthal2024single}.
The cyclicity $\chi$ is estimated as $\chi = \gamma_0 t_{\text{sat}} / 2$ where $\gamma_0/2\pi$ is the lifetime-limited linewidth (full-width at half maximum (FWHM)) of the \gev center's optical transition and $t_{\text{sat}}$ is the mean spin pumping time at saturation laser power $p_\text{sat}$.
As laser power $p$ is increased, the mean spin pumping time $t_{\text{init}}$ generally decreases due to faster driving of the optical transitions.
At $p_\text{sat}$, the optical transition rate is saturated and $t_{\text{init}}$ plateaus.
This behavior is shown in \autoref{fig:fig9_init} and we estimate $t_{\text{init}} \approx 81\,\text{\textmu s}$.
\autoref{fig:fig9_fwhm} shows the behavior of $\gamma$ with optical power $p$, allowing for the determination of $\gamma_0/2\pi\approx44\,\text{MHz}$.
This allows for the estimation of the cyclicity $\chi \approx 11000$.
We can further estimate the collection efficiency of the system as $\sim0.09\%$ by comparing the cyclicity with the actual number of photons collected during readouts.
Note that this collection efficiency estimate includes all losses along the detection path and the detector efficiency as well.

\section{Correlation Spectroscopy Measurement Time}

We highlight that while the measurement in the inset of Figure 2\,(c) in the main text consists of 461 shots for each of the 8192 $\tau$ steps and took over 44\,h of measurement time, most of which was the combined laser length and waiting time of 20\,ms to minimize heating of the system, we have performed measurements with an order of magnitude fewer shots and still observed the same features. \\
Measurements with shorter waiting times between shots showed substantially lower signal-to-noise ratios, likely due to phonon-induced dephasing of the \gev center during the application of mw pulses during the DD blocks. \\
With a lower-loss transmission line, we can expect to use a 5\,ms laser length or less, resulting in at least a fourfold reduction in measurement time.

\section{Correlation Spectroscopy Spurious Harmonics Suppression}

\begin{figure}
    \centering
    \includegraphics{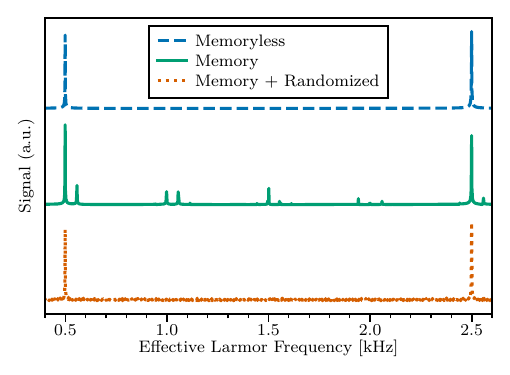}
        \caption{
        Simulation showing that spurious harmonics can arise in correlation spectroscopy measurements.
        The simulation only considers  \nucc$_\text{B}$ for simplicity.
        The simulation shows that randomizing the order in which $\tau$ is swept is sufficient to mitigate spurious harmonics.
        Note that vertical offsets are added for better visualization.
        }
    \label{fig:fig7_1d}
\end{figure}

The simulation in \autoref{fig:fig7_1d} only considers \nucc$_\text{B}$ for simplicity.
The `memoryless' case is the ideal case where the \nuc spin is completely reset to a maximally mixed state after every measured $\tau$ step.
The `memory' case is where only the \gev center's spin is reset and the \nuc spin is not, leading to unwanted harmonic signals.
The `memory' case with randomization is where the order in which the $\tau$ are measured is randomized and subsequently sorted prior to the Fourier transform and shows that the spurious harmonics can be mitigated.
This works by converting the spurious correlations between consecutive measured $\tau$ steps into uncorrelated noise when the measured data is sorted prior to the Fourier transform.
We note that adding a waiting time significantly longer than \Ttwostar[n] is also a possible strategy to mitigate these spurious harmonics which is what we are currently doing with the long laser pulse which we use to mitigate heating effects.
We therefore cannot easily verify these spurious harmonics experimentally because we need a waiting time on the order of \Ttwostar[n] which would cause heating issues in our current system.
Nevertheless, we do the $\tau$ randomization for all correlation spectroscopy experiments in this work to further enhance our confidence in small signals being accurately attributed to \nuc spins.
\begin{figure}
    \centering
    \includegraphics{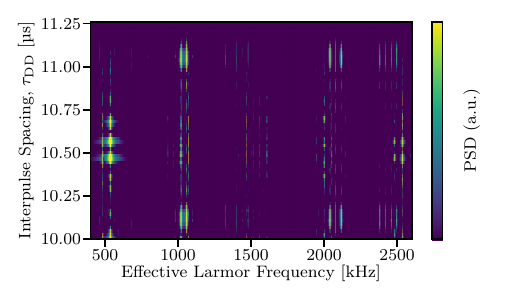}
        \caption{
        Simulation showing spurious harmonics in a two-dimensional correlation spectroscopy measurement.
        The simulation considers both \nucc$_\text{A}$ and \nucc$_\text{B}$ and highlights the significant challenge that these harmonics can cause in the reliable identification of single \nuc spins.
        }
    \label{fig:fig7_2d_harmonics}
\end{figure}

\begin{figure}
    \centering
    \includegraphics{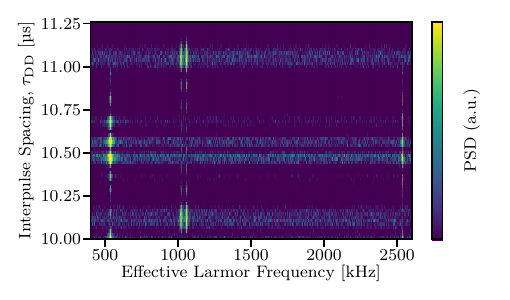}
        \caption{
        Simulation showing how $\tau$ randomization works in a two-dimensional correlation spectroscopy measurement to mitigate spurious harmonics.
        The simulation considers both \nucc$_\text{A}$ and \nucc$_\text{B}$ and only 4 distinct effective Larmor frequencies are visible as expected.
        }
    \label{fig:fig7_2d_randomised}
\end{figure}

\section{$\text{C}_n\text{NOT}_e$ Rabi Frequency Optimization for Nuclear Single-Shot Readout}
The \Ttwostar[e] of our system limits our lowest $\text{C}_n\text{NOT}_e$ Rabi frequency to around 350 kHz.
However, we have observed spectral diffusion of the $\nu_1$ and $\nu_2$ optically detected magnetic resonance (ODMR) lines~\cite{grimm2025coherent} of $\pm\,150$ kHz resulting in significant detuning of the Rabi drive and a reduced SSR fidelity.
While an arbitrarily higher Rabi frequency could compensate for the spectral diffusion-induced detuning and also be more robust to dephasing due to a shorter pulse duration, the associated power broadening would generally reduce the (frequency) selectivity of the pulse and correspondingly reduce the nuclear SSR fidelity.
This can be seen in the expression for generalized Rabi frequency
\begin{equation}
    \Omega_{\mathrm{eff}} = \sqrt{\Omega^2 + \Delta^2}
    \label{eq:gen_rabi_freq}
\end{equation}
and generalized Rabi amplitude
\begin{equation}
    a = \frac{\Omega^2}{\Omega^2 + \Delta^2},
    \label{eq:gen_rabi_amp}
\end{equation}
where $\Omega$ is the (on-resonance) Rabi frequency
and $\Delta$ is the detuning of the driving frequency relative to the resonance frequency.
While a lower $\Omega$ such that $\frac{\Omega}{\Delta} \ll 1$
would minimize $a$, it would mandate a pulse length exceeding \Ttwostar[e] and also be significantly affected by detuning due to spectral diffusion, resulting in poor gate fidelity.
We can instead choose an $\Omega$ that results in an integer multiple of detuned $2\pi$ Rabi oscillations
on the off-resonant transition when the resonant transition experiences a $\pi$ Rabi oscillation i.e.
\begin{equation}
    \frac{1}{2\,\Omega} = \frac{m}{\sqrt{\Omega^2 + \Delta^2}}
    \label{eq:ssr_detuning}
\end{equation}
where $m \in \mathbb{Z^+}$.
Rearranging, we get
\begin{equation}
    \Omega = \frac{\Delta}{\sqrt{4m^2-1}}.
    \label{eq:ssr_rabi_target}
\end{equation}

At $m=2$ theoretically, we obtain an optimal Rabi frequency of $2\pi \cdot 743\,$kHz, slightly lower than the $\Omega=2\pi \cdot 800\,$kHz used in the experiment. To investigate the spin-flip fidelity and the back action of the gate on the nuclear spin, we simulate a closed two-body system according to Eq.~(1) of the main text with $\Delta=A_{zz}$ (i.e., the mw drive is on resonance with the electron spin flip), $\theta=0$ and $\Omega=2\pi \cdot 800\,$kHz for a total time of $\pi/\Omega$. If we set $A_{zx}=0$, we obtain a perfect electron spin flip for $\ket{\downarrow_e \downarrow_n}$, with no back action on the nuclear spin. If we start in $\ket{\downarrow_e \uparrow_n}$, we obtain an electron spin-flip probability of about $1.6\,\%$ and no back-action on the nuclear spin ($<10^{-6}$). If we now turn on $A_{zx}$ to its true value, we obtain a back action of the readout gate on the nuclear spin. The reason for this is that in the presence of a perpendicular hyperfine interaction, the eigenstates of the static Hamiltonian are slightly tilted with respect to the $z$-axes~\cite{grimm2025coherent} and thus the otherwise forbidden double-transitions (in this case mainly from $\ket{\downarrow_e\downarrow_n}$ to $\ket{\uparrow_e\uparrow_n}$) are allowed.
For the initial state $\ket{\downarrow_e \downarrow_n}$ the electron spin-flip probability slightly drops to about $99\,\%$, while the nuclear spin flips at almost $12\,\%$. For the initial state $\ket{\downarrow_e \uparrow_n}$, the drive is off-resonant and thus also with $A_{zx}$ present, the electron spin-flip probability is almost unchanged at about $1.8\,\%$, while the back action on the nuclear spin leads to a flipping probability of about $1\,\%$.

\section{Readout Threshold Optimization for Nuclear Single-Shot Readout}

\begin{figure}
    \centering
    \includegraphics{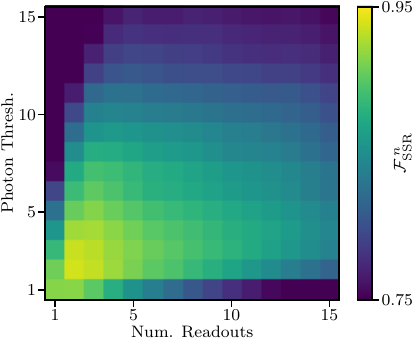}
        \caption{
        Heatmap showing how the single-shot readout fidelity of \nucc$_\text{A}$ varies with photon threshold and number of readouts for the measured optimal Rabi frequency of $\Omega = (2\pi)\,800\,\text{kHz}$.
        }
    \label{fig:fig8}
\end{figure}

In the main text, Figures 3(a) and 3(b) calculate $\mathcal{F}_{\text{SSR}}^{\,n}$ using two consecutive readouts with a total accumulated photon threshold of two.
These optimal readout threshold parameters were found by performing nuclear SSR with a fixed Rabi frequency of the readout $\text{C}_{n}\text{NOT}_e$ gate of $800\,\text{kHz}$ and calculating $\mathcal{F}_{\text{SSR}}^{\,n}$ for various permutations of the number of consecutive readouts and total accumulated photons (i.e. photon threshold).
The results of this measurement are shown in \autoref{fig:fig8}.

\newpage
\bibliographystyle{ieeetr}
\bibliography{references}